\documentclass[%
 reprint,%
 amssymb, amsmath,%
 aip,cha,%
]{revtex4-1}
\usepackage[columnwise]{lineno}
\usepackage{docs}%
\usepackage{bm}%
\usepackage[colorlinks=true,linkcolor=blue]{hyperref}%
\expandafter\ifx\csname package@font\endcsname\relax\else
 \expandafter\expandafter
 \expandafter\usepackage
 \expandafter\expandafter
 \expandafter{\csname package@font\endcsname}%
\fi
\hypersetup{
	colorlinks=true,
	linkcolor=blue,
	filecolor=magenta,}
\usepackage{graphicx}
\usepackage{lipsum}

\begin{document}

\title{Magnetism at the interface of non-magnetic Cu and C$_{60}$} %

\author{Purbasha Sharangi}%
\affiliation{Laboratory for Nanomagnetism and Magnetic Materials (LNMM), School of Physical Sciences, National Institute of Science Education and Research (NISER), HBNI, P.O.- Jatni, 752050, India}%
\author{Pierluigi Gargiani}%
\affiliation{Alba Synchrotron Light Source, E-08290 Barcelona, Spain}%
\author{Manuel Valvidares}%
\affiliation{Alba Synchrotron Light Source, E-08290 Barcelona, Spain}%
\author{Subhankar Bedanta}%
\email{sbedanta@niser.ac.in}
\affiliation{Laboratory for Nanomagnetism and Magnetic Materials (LNMM), School of Physical Sciences, National Institute of Science Education and Research (NISER), HBNI, P.O.- Jatni, 752050, India}%
\date{September 2020}%


\begin{abstract}
The signature of magnetism without a ferromagnet in a non-magnetic heterostructure is novel as well as fascinating from fundamental research point of view. It has been shown by Al'Mari $et$ $al.$ that magnetism can be induced at the interface of Cu/C$_{60}$ due to change in density of states. However, the quantification of such interfacial magnetic moment has not been performed yet. In order to quantify the induced magnetic moment in Cu, we have performed X-ray magnetic circular dichroism (XMCD) measurements on Cu/C$_{60}$ multilayers. We have observed room temperature ferromagnetism in Cu/C$_{60}$ stack. Further XMCD measurements show that $\sim$ 0.01 $\mu_B$ /atom magnetic moment has been induced in Cu at the Cu/C$_{60}$ interface.

\end{abstract}

\maketitle
 Organic semiconductors (OSC) are potential candidates for spintronics based applications due to various reasons such as low spin orbit coupling (light weight element), mechanical flexibility, and versatility of material synthesis \cite{sun2014first,dediu2009spin,atodiresei2010design,barraud2010unravelling}. Buckminsterfullerene (C$_{60}$) has drawn immense research interest in organic spintronics due to its structural simplicity, robustness, and high electron affinity. Large spin dependent transport length and large spin relaxation time ($>$ 1$\mu$s) have been observed using C$_{60}$ as a spacer layer in between two ferromagnet (FM) \cite{nguyen2013spin}. Due to the absence of hydrogen and associated hyperfine interaction, C$_{60}$ exhibits less spin scattering. Hence, it has large spin dependent transport length in comparison to conventional inorganic semiconductors. Spin polarized charge transfer occurs at the FM-OSC interface due to orbital hybridization leading to the modification of density of states \cite{sanvito2010molecular,djeghloul2013direct}. An induced moment of 1.2 $\mu_B$ per cage of C$_{60}$ and suppression of magnetic moment in Co has been observed for Co/C$_{60}$ multilayers by X-ray magnetic circular dichroism (XMCD) and polarized neutron reflectivity (PNR) measurements \cite{moorsom2014spin}. It has been reported that C$_{60}$ monolayers on Fe (001) reveal magnetic polarization of C$_{60}$ due to the hybridization of C$_{60}$ and Fe orbitals \cite{tran2011hybridization}. Similarly, for Fe/C$_{60}$ system a suppression of magnetic moment of Fe and an induced magnetic moment of 2.95 $\mu_B$ per cage of C$_{60}$ have been observed \cite{mallik2018effect}. However, emergence of ferromagnetism at room temperature without incorporating any FM layer in the sample stack is another direction of organic spintronics. It has been reported that it is possible to alter the electronic states of non-ferromagnetic materials (Cu, Mn, Sc, Pt) to overcome the Stoner criterion and make them ferromagnetic at room temperature \cite{al2015beating,felici2005x}. Charge transfer and interface reconstruction have been observed at the Cu/C$_{60}$ interface. This results in modifications of the density of states (DOS) of the Cu layer and a band splitting, which leads to magnetic ordering \cite{al2015beating}. Density functional theory calculation and high-resolution angle-resolved photoemission spectroscopy revealed that the modification in the electronic structure occurs at the interface between a highly ordered C$_{60}$ monolayer and Cu (111) surface \cite{stengel2003adatom,shoup2020structural}. In this context it is desired to quantify the magnetic moment at the interface of such non-FM/OSC layers.  In this paper we have studied the magnetic properties of Cu/C$_{60}$ heterostructure and quantified the magnetic moment induced in Cu using XMCD sum rules.

 \begin{figure}[h!]
 	\centering
 	\includegraphics[width=1.0\linewidth]{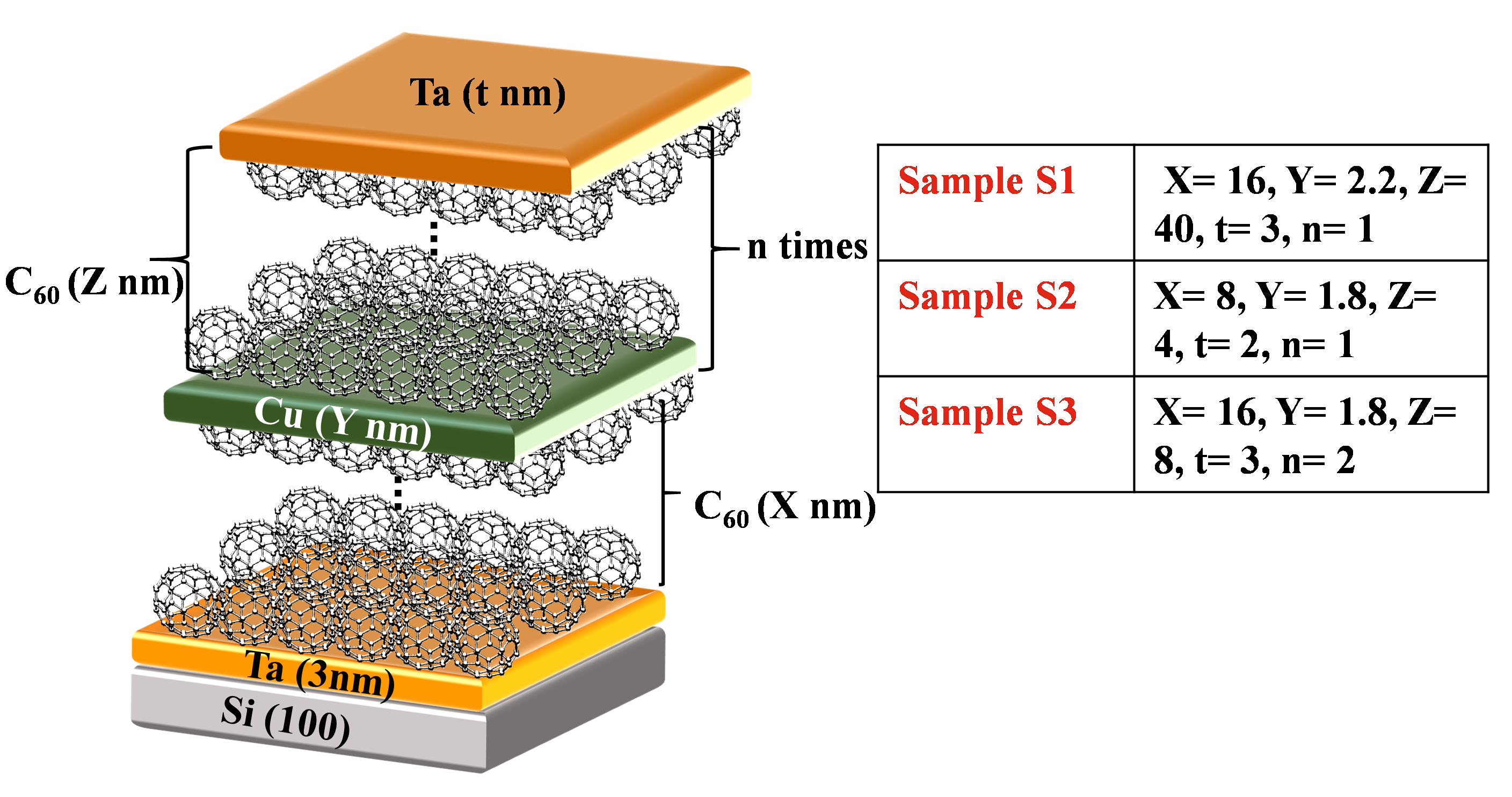}
 	\caption{Schematic (not to be scaled) of the prepared samples structure.}
 	\label{fig:fig1}
 \end{figure}
 
 \begin{figure*}
 	\centering
 	\includegraphics[width=1.0\linewidth]{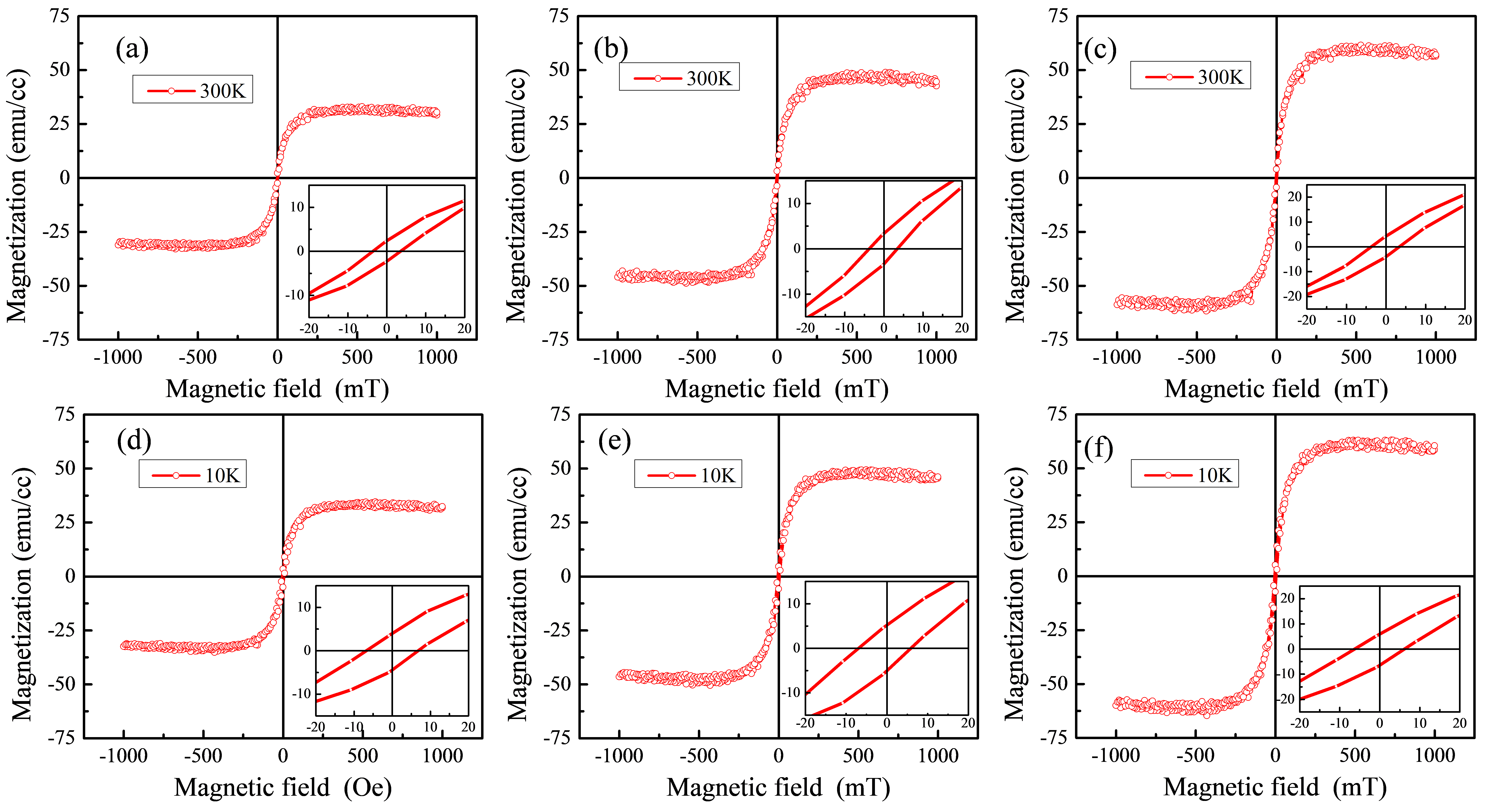}
 	\caption{Room temperature (300K) $M$-$H$ loops using SQUID magnetometer are shown for sample S1 (a), sample S2 (b), sample S3 (c) and  10 K for sample S1 (d), sample S2 (e) and sample S3 (f).}
 	\label{fig:fig2}
 \end{figure*}
 
We have prepared multilayers of Cu/C$_{60}$ on Si (100) substrate using DC magnetron sputtering and thermal evaporation techniques for Cu and C$_{60}$, respectively, in a multi-deposition HV chamber manufactured by Mantis Deposition Ltd., UK. The base pressure of the deposition chamber was better than 5$\times$10$^{-8}$ mbar. All the Cu and C$_{60}$ layers have been deposited without breaking the vacuum. The deposition pressure was 5$\times$10$^{-3}$ mbar for Cu and 1$\times$10$^{-7}$ mbar for C$_{60}$ evaporation. The Cu and C$_{60}$ layers were deposited at a rate of 0.1 and $\sim$ 0.1 – 0.15 \AA/s, respectively. For better growth of Cu, a 5 nm thick Ta layer was taken as a seed layer. The schematic (not to be scaled) of the sample structure is shown in figure 1. The sample structure is the following: Si/Ta(3)/C$_{60}$(X)/[Cu(Y)/C$_{60}$(Z)]$\times$n/Ta(t), where X, Y, Z and t are in nm and the values are referred in figure 1. 
To prevent oxidation, a capping layer of Ta has been deposited on top of C$_{60}$.

To estimate the interdiffusion of the Cu/C$_{60}$ and C$_{60}$/Cu, we have performed X-ray reflectivity (XRR) measurements with the X-ray diffractometer (XRD) manufactured by Rigaku. We have carried out in-plane field dependent magnetic measurements ($M$-$H$) by superconducting quantum interferences device (SQUID) magnetometry (MPMS3) manufactured by Quantum Design, USA. The magnetic field was applied along the film plane. XMCD is the perfect tool to determine the localized magnetization and quantify the element specific magnetic moment. The XMCD measurements were performed at BOREAS beamline at Alba Synchrotron Light Source, E-08290 Barcelona, Spain \cite{barla2016design}. In order to excite the core electron, circularly polarized X-rays were directed onto the sample with an energy of 80-1500 eV and maximum resolution of $\Delta E/E = 10^{‐4} $. The electrons released from the sample via this process were collected as a drain current in a total electron yield (TEY) mode. To saturate the sample, $\pm$ 6 T magnetic field was applied collinear to the impinging X-rays. The energy was calibrated at the beginning of the experiment with the known CoO reference. All the measurements were performed in a UHV condition with a base pressure better than 2$\times$ 10$^{-10}$  mbar and at a sample temperature of 1.7 K.

Interface plays an important role to induce magnetism at Cu/C$_{60}$ interface. Generally the interdiffusion of metal/OSC interface is higher than that of OSC/metal. From XRR fit (shown in supplementary figure S1) we found that interdiffusion is present at both the Cu and C$_{60}$ interfaces. The thickness of the interdiffused layers are 0.58 and 0.53 nm for the Cu/C$_{60}$ and C$_{60}$/Cu interfaces, respectively. 
Figure 2 shows the hysteresis loops measured by SQUID magnetometer at 10 and 300 K. It is observed that all the samples exhibit ferromagnetism even when no ferromagnetic element is present in the samples. The coercivities ($H_{C}$) at 10 K for samples S1, S2 and S3 are 6.40, 6.55 and 6.50 mT, respectively. Further, at 300 K, the $H_{C}$ values for samples S1, S2 and S3 are 3.60, 3.80 and 3.75 mT, respectively. It has been observed that magnetization increases with number of Cu/C$_{60}$ interfaces, which is in agreement with the previous report \cite{al2015beating}. Magnetization also depends on the thickness of the Cu layer. From the hysteresis loops we have observed that magnetization is slightly higher for samples S2 and S3, where the thickness of Cu is 1.8 nm.

\begin{figure*}
	\centering
	\includegraphics[width=1.0\linewidth]{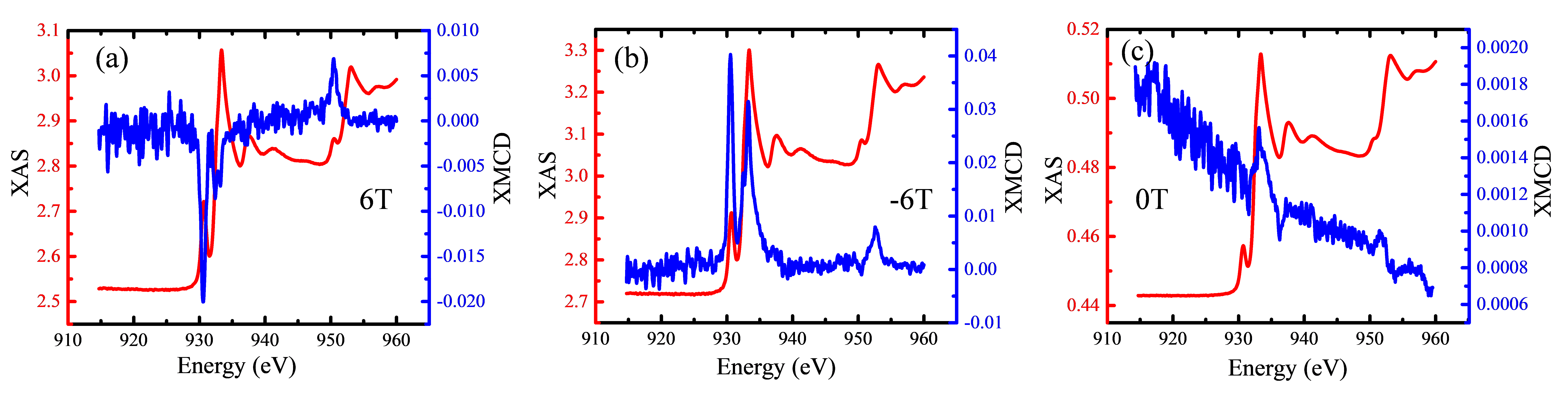}
	\caption{XAS and XMCD spectra of the sample S2 measured at (a) 6 T, (b) -6 T and (c) 0 T magnetic field at Cu $L_{2,3}$ edges. All the measurements were performed at 1.7 K.}
	\label{fig:fig3}
\end{figure*}

\begin{figure*}
	\centering
	\includegraphics[width=0.8\linewidth]{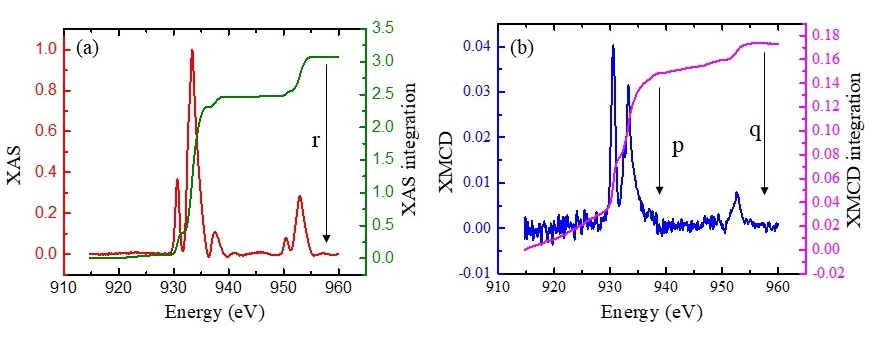}
	\caption{Cu $L_{2,3}$ XAS (a) and XMCD (b) spectra and their integrations calculated from the spectra are shown for the sample S2 at -6 T. The green solid line is the integral of the XAS after subtracting two-step-like function from XAS spectra. The red solid line represents the spectra after subtracting a two-step function from XAS spectra. The p, q and r are the three integrals needed in the sum-rule analysis.}
	\label{fig:fig4}
\end{figure*}

Magnetic moment is observed in the samples probably due to the charge transfer and the molecular coupling between the metal (Cu) and C. C$_{60}$ induced interface reconstruction have been observed for C$_{60}$/Au (110) \cite{atodiresei2010design}, C$_{60}$/Pt (111) \cite{felici2005x}, C$_{60}$/Al (111) \cite{stengel2003adatom}, C$_{60}$/Ag (100) \cite{pai2003ordering}, and even for C$_{60}$/Ag (111) \cite{li2009surface} and C$_{60}$/Cu (111) \cite{pai2004structural} systems. Reconstructed C$_{60}$/Cu (111) interface has a  1-3 electron transfer per C$_{60}$ cage whereas, an unreconstructed one receives a much smaller amount ($<$ 0.8e-) \cite{tsuei1997photoemission,wang2004rotation}. The origin of the charge state of C$_{60}^{-3}$ to a reconstructed interface is due to (4 $\times$ 4) 7-atom vacancy holes in the surface \cite{pai2010optimal}.The possible reason of this induced magnetic moment is hybridization between $d_{Cu}$ and $p_{C_{60}}$ orbitals \cite{al2015beating,al2017emergent,tamai2008electronic,raman2015materials}. Cu has the ability to transfer up to 3 electrons per C$_{60}$ cage due to the high electron affinity of C, which modifies the density of state of Cu \cite{al2015beating,cho2007origin} . Further, the modified density of states of Cu satisfies the Stoner Criteria  and exhibits ferromagnetism  \cite{al2015beating}. 

\begin{figure*}
	\centering
	\includegraphics[width=0.8\linewidth]{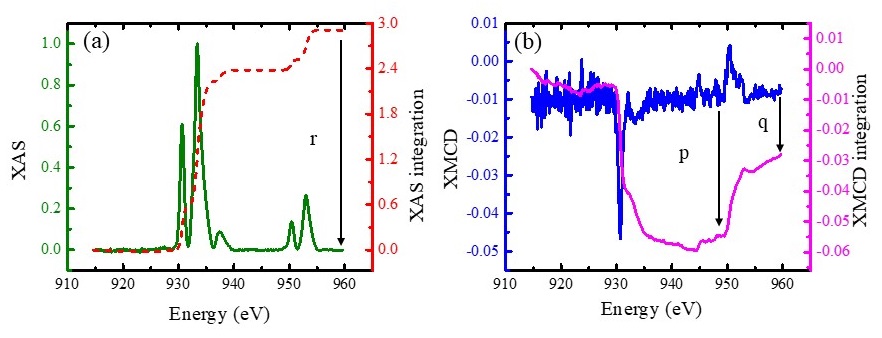}
	\caption{Cu $L_{2,3}$ XAS (a) and XMCD (b) spectra and their integrations calculated from the spectra are shown for the sample S3 at 6 T. The red dotted line is the integral of the XAS after subtracting two-step-like function from XAS spectra. The green solid line represents the spectra after subtracting a two-step function from XAS spectra. The p, q and r are the three integrals needed in the sum-rule analysis.}
	\label{fig:fig5}
\end{figure*}

XMCD determines the difference between two X-ray absorption spectra (XAS) recorded under a magnetic field, one taken with right circularly polarized x-rays and the other one with left circularly polarized x-rays. Analysis of the XMCD spectrum gives  information about the electronic and magnetic properties of the atoms, such as orbital and spin angular momentum. Using magneto-optical sum rules one can obtain the ground state expectation values of the orbital ($L_{z}$) and the spin ($S_{z}$) angular momentum \cite{thole1992x,carra1993x,chen1995experimental,stohr1999exploring}. Hence, XMCD is an efficient experimental tool to study element specific magnetic properties. For 3$d$ transition metals the experiments are performed at the $L_{2,3}$ absorpion edges (2$p$ -$>$ 3$d$ transition), since the magnetic moment is mostly carried out by the 3$d$ electrons \cite{thole19853d}.

 Figure 3 shows the XAS and XMCD spectra of the sample S2 at (a) 6 T, (b) -6 T, and (c) 0 T magnetic field, respectively. The same for sample S3 are shown in supplementary (figure S2). The sign of the dichroism was changed when a negative magnetic field was applied, confirming that the measured signal was not due to spurious experimental effects. To compare the XMCD intensities we have normalized XAS spectra at $L_3$ edge. Integration of XMCD signals at $L_2$ and $L_3$ edges lead to the orbital and spin magnetic moment under the applications of magneto-optical sum rules \cite{thole1992x,chen1995experimental,stohr1999exploring,thole19853d}. Cu $L_{2,3}$ edges spectra are observed at 933.4 eV and 953.06 eV, which correspond to the transition from 2$p$ to 3$d$ state. A very small difference between the $L_2$ and $L_3$ edge XMCD intensities offers a small anisotropic orbital magnetic moment in Cu  \cite{okabayashi2017induced}. We have observed a XMCD signal from the pre-peak which corresponds to Cu$_{2}$O (930.8 eV) similar to the results reported by Ma’Mari $et$ $al$. \cite{al2015beating,van1992oxidation,greiner2015oxidation}. However, we have also observed a significant XMCD signal from the Cu $L_2$ and  $L_3$ peaks. It has been shown by Ma'Mari $et$ $al$. that on introduction of Al or Al$_{2}$O$_{3}$ between Cu and C$_{60}$ layers resulted in vanishing of magnetization. This indicates that the interface between Cu/C$_{60}$ plays a big role in inducing the magnetism at Cu \cite{al2015beating}. The Cu atoms in Co/Cu multilayers exhibit induced magnetism due to exchange coupling between $d$ electrons of Cu and Co layers \cite{samant1994induced}. Orbital and spin angular momentum are calculated using the following sum rule formula \cite{thole1992x,chen1995experimental,stohr1999exploring,thole19853d}:
\begin{equation}
	m_{orb}=    \frac{-4q (10-n_{3d})}{3r}
\end{equation}

\begin{equation}
	m_{spin effective}=     \frac{-(6p-4q)(10-n_{3d})}{r}
\end{equation}

where, $q= \int_{L_2+L_3}(\mu_+ - \mu_-)dE$, $p= \int_{L_3}(\mu_+ - \mu_-)dE$,       
$r= \int_{L_2+L_3}(\mu_+ + \mu_-)dE$, $m_{orb}$ and $m_{spineffective}$   are the orbital and spin magnetic momentum in units of $\mu_B$/atom, respectively. $n_{3d}$ is the 3$d$ electron occupation number of the specified transition metal. $L_{2}$ and $L_{3}$ denote the integration range. The ratio of orbital to spin magnetic moments has been calculated using the following equation \cite{stohr1999exploring}:

\begin{equation}
	\frac{m_{orb}}{m_{spin effective}}=     \frac{-2q}{9p-6q}
\end{equation}

Figure 4 shows $L_{2,3}$ edge XMCD, summed XAS spectra and the integrations calculated from the spectra for sample S2 at -6 T. The same for sample S3 at 6 T field has been shown in figure 5. We have subtracted a two-step function from XAS spectra before the integration to remove all the contribution which does not come from the 2$p$ - 3$d$ transition. The integral for the whole $L_{3} + L_{2}$ range (q value) and  for the $L_{3}$ edge (p value) can be precisely determined from the integrated spectrum,  which are shown in figure 4 and 5. The r value corresponds to the XAS integral in the individual sum rule calculation. Using XMCD sum rules we have calculated the orbital and spin magnetic moments for Cu \cite{thole1992x,chen1995experimental,stohr1999exploring,thole19853d}. The spin magnetic moment of Cu for sample S2 and S3 are 0.0078$\pm$0.0019 and 0.0116$\pm$0.0032 $\mu_B$/atom, respectively. We have chosen the hole numbers $n_{Cu}$=0.44 \cite{okabayashi2017induced}. The sum rule analysis for sample S2 at 6 T field also yielded 0.0071$\mu_B$/atom  magnetic moment induced in Cu (shown in supplementary figure S3). Although paramagnetism in Cu has been reported previously \cite{ebert2003field,yaouanc2004comment}, our SQUID magnetometry and XMCD signal confirm that the magnetism is coming from the Cu/C$_{60}$ interface is not due to the metallic state of Cu.

In conclusion, we have investigated the induced magnetic moment in Cu/C$_{60}$ interface via SQUID magnetometry and XMCD. Due to the charge transfer at the reconstructed Cu/C$_{60}$ interface, the density of state of Cu is modified and exhibits a magnetic moment of $\sim$0.01$\mu_B$/atom. Future work may bring new insights to (i) which other non-magnetic metals can also exhibit ferromagnetism in such NM/OSC heterostructures, (ii) why Cu exhibits such FM only at ultrathin limit; (iii) exploration with other organic materials to exhibit similar physical phenomena etc. The answers to these questions will have significant importance in the field of organic spintronics.

This work is financially supported by Department of Atomic Energy, and Department of Science and Technology - Science and Engineering Research Board, Govt. of India (DST/EMR/2016/007725). The authors thank NFFA-Europe for funding the XMCD measurement (proposal ID488). The authors would also like to thank Dr. Srijani Mallik for discussion and help during sample preparations.

\bibliography{references}
\end{document}